\documentclass[sigconf]{acmart}
\usepackage{caption}
\usepackage{graphicx}
\usepackage{float} 
\usepackage{subcaption}
\usepackage{enumitem}
\usepackage{amsmath}
\usepackage{algorithmic}
\usepackage{algorithm}
\usepackage{array}
\usepackage{tikz}
\newcommand*\circled[1]{\tikz[baseline=(char.base)]{
		\node[shape=circle, fill=black, inner sep=0.5pt, text=white] (char) {#1};}}

\renewcommand\footnotetextcopyrightpermission[1]{}

\begin{document}
	
\title{SoberDSE: Sample-Efficient Design Space Exploration via Learning-Based Algorithm Selection}

\author{Lei Xu, Shanshan Wang, and Chenglong Xiao$^\dagger$}
\affiliation{
\institution{Department of Computer Science, Shantou University}
\city{Shantou}
\country{China}}
\email{{24lxu, sswang, chlxiao}@stu.edu.cn}
\thanks{*Corresponding author: Chenglong Xiao (chlxiao@stu.edu.cn). This work is partially sponsored by Guangdong Basic and Applied Basic Research Foundation (2022A1515110712, 2025A1515010272, 2023A1515010077) and the Scientific Research Project of Colleges and Universities in Guangdong Province (2021ZDZX1027).}

\begin{abstract}
High-Level Synthesis (HLS) is a pivotal electronic design automation (EDA) technology that enables the generation of hardware circuits from high-level language descriptions. A critical step in HLS is Design Space Exploration (DSE), which seeks to identify high-quality hardware architectures under given constraints. However, the enormous size of the design space makes DSE computationally prohibitive. Although numerous algorithms have been proposed to accelerate DSE, our extensive experimental studies reveal that no single algorithm consistently achieves Pareto dominance across all problem instances. Consequently, the inability of any single algorithm to dominate all benchmarks necessitates an automated selection mechanism to identify the best-performing DSE algorithm for each specific case. To address this challenge, we propose the SoberDSE framework, which recommends suitable algorithm based on benchmark characteristics. Experimental results demonstrate that our SoberDSE framework significantly outperforms state-of-the-art heuristic-based DSE algorithms by up to 5.7 $\times$ and state-of-the-art learning-based DSE methods by up to 4.2 $\times$. Furthermore, compared to conventional classification models, SoberDSE delivers superior accuracy in small-sample learning scenarios, with an average enhancement of 35.57\%. Code and models are available at \url{https://anonymous.4open.science/r/Sober-4377}.
\end{abstract}
%
%


\maketitle

\section{INTRODUCTION}
High-Level Synthesis (HLS) tools provide high-level, abstract interfaces that assist hardware designers in developing hardware designs while avoiding low-level implementation complexities \cite{Schafer2019}. By enabling programming at behavioral and register-transfer levels, HLS significantly reduces the expertise barrier for hardware design. These tools also offer a comprehensive set of synthesis parameters, including \verb|#pragma HLS unroll|, \verb|#pragma HLS pipeline|, and \verb|#pragma HLS array_partition|, to guide optimization across power, performance, and area (PPA) metrics. However, identifying the optimal parameter combination through Design Space Exploration (DSE) often requires a considerable time investment, as the configuration space grows exponentially with the number of pragma directives. Additionally, each high-level design iteration can take anywhere from minutes to days to simulate using commercial HLS tools such as Vivado HLS or Vitis, which makes exhaustive exploration practically infeasible for complex designs. This computational bottleneck underscores the need for intelligent DSE methodologies that can efficiently navigate large design spaces without compromising solution quality.

\begin{figure}[t]
	\centering
	\begin{subfigure}{0.98\linewidth}
		\centering
		\includegraphics[width=0.99\linewidth]{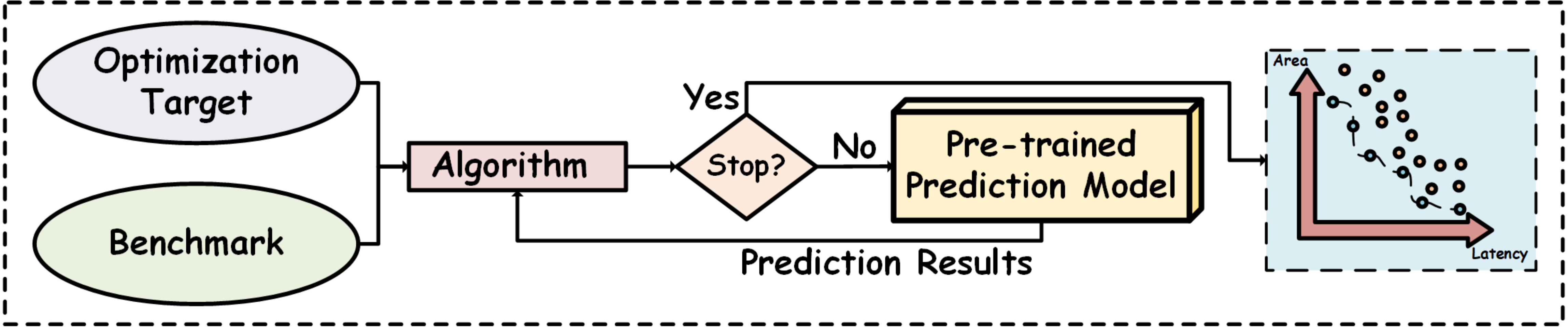}
		\caption{Framework for DSE using predictive models.}
		\label{fig:HLSDSE}
	\end{subfigure}

	\begin{subfigure}{0.98\linewidth}
		\centering
		\includegraphics[width=0.99\linewidth]{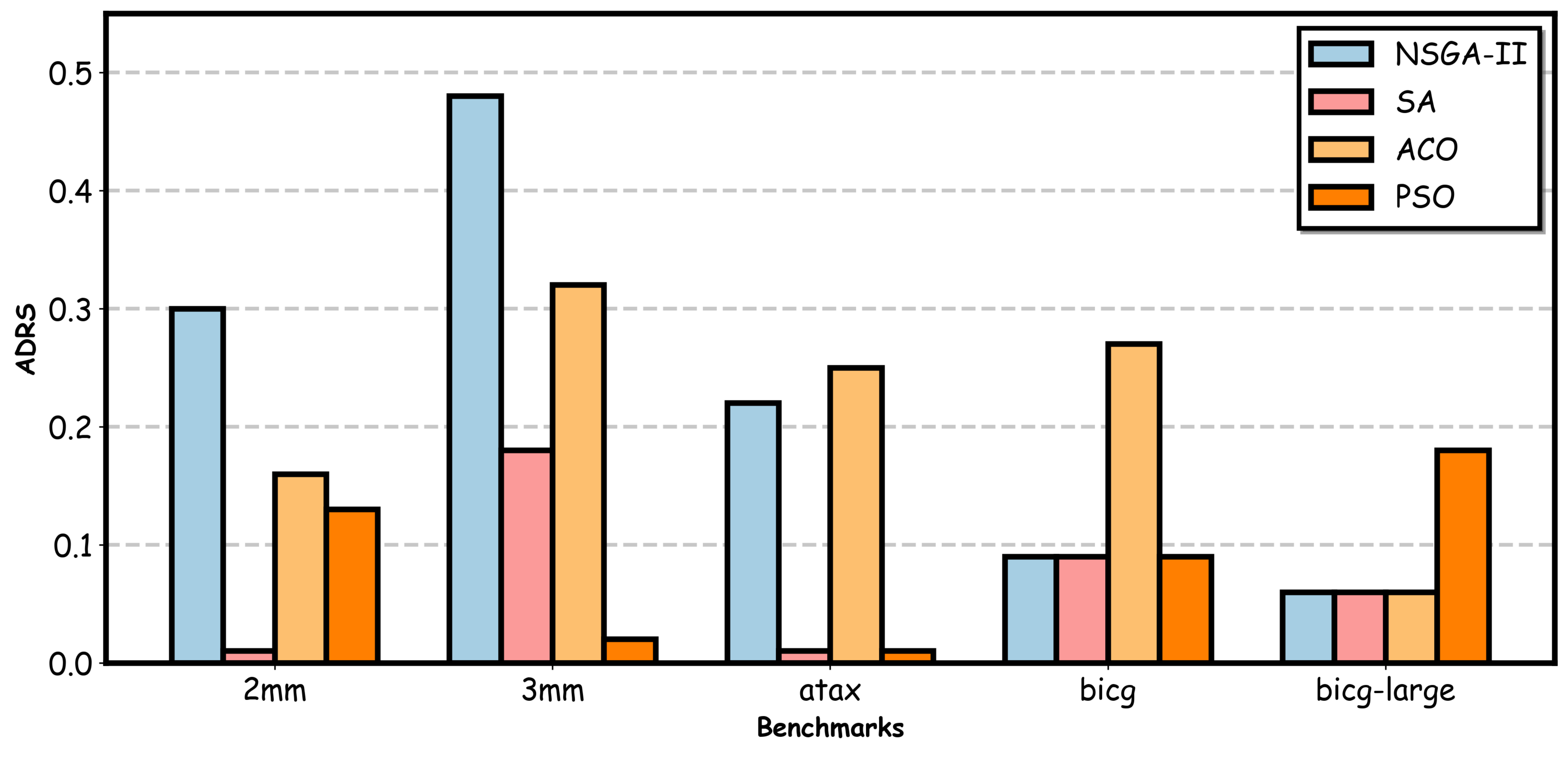}
		\caption{Actual ADRS obtained by heuristic techniques on five benchmarks.}
		\label{fig:mt}
		\vspace{-2mm}
	\end{subfigure}
	\centering
	\caption{Motivation of Our Proposed \textit{SoberDSE}.}
	\label{fig:mo}
	\vspace{-3mm}
\end{figure}

To address these challenges, a promising approach involves converting high-level designs into Control and Data Flow Graphs (CDFGs) and leveraging graph representation learning techniques to extract structural features that capture both control and data dependencies inherent in the original code. This enables rapid prediction of key performance metrics, such as latency and resource utilization, without relying on time-consuming HLS tool simulations. Furthermore, heuristic methods \cite{Schafer2017, Schafer2009, Schafer2016, Zhang2015} are employed to efficiently search for near-optimal synthesis parameters within the vast design space, achieving a dynamic balance among Power, Performance, and Area (PPA) objectives. For example, as shown in Figure \ref{fig:HLSDSE}, the DSE algorithm conducts design space exploration based on specific optimization objectives and benchmark characteristics. The process begins by checking whether termination conditions, such as maximum iterations or convergence criteria, are satisfied. If met, the Pareto approximate optimal solution set is output. Otherwise, the pre-trained graph neural network (GNN) model provides accurate predictions to guide subsequent exploration iterations, creating an efficient feedback loop that progressively refines the solution quality while significantly reducing exploration time compared to conventional methods.

However, as established by the No Free Lunch (NFL) theorem \cite{Wolpert2002}, there is no single algorithm capable of achieving optimal performance across all instances within a given class of optimization problems. This fundamental principle implies that any improvement an algorithm exhibits on one set of problems will inevitably be counterbalanced by degraded performance on another set. DSE, being a multi-objective optimization task involving trade-offs among conflicting metrics such as power, latency, and resource utilization, has been tackled in prior studies using a variety of methods, including heuristic algorithms and reinforcement learning (RL) based approaches. Nevertheless, consistent with the NFL theorem, none of these algorithms has been shown to consistently outperform all others across  all instances. As illustrated in Figure \ref{fig:mt}, we conduct DSE using four heuristic algorithms non-dominated sorting genetic algorithm II (NSGA-II) \cite{Schafer2017}, simulated annealing algorithm (SA) \cite{Schafer2009}, ant colony optimization (ACO) \cite{Schafer2016}, and Particle Swarm Optimization (PSO) \cite{Zhang2015} on benchmarks including \textit{2mm}, \textit{3mm}, \textit{atax}. \textit{bicg} and \textit{bicg-large} with the Average Distance to Reference Set (ADRS) as the performance metric (lower values indicate better performance). The results show that SA performs best on 2mm and atax, while different algorithms excel on the remaining benchmarks. These experimental findings confirm that no DSE algorithm performs optimally across all benchmarks, underscoring the necessity of an automated system to recommend the most suitable algorithm on a case-by-case basis.

To address this problem, we propose SoberDSE, a framework that integrates the concept of algorithm selection to automatically recommend suitable algorithms for each benchmark based on the characteristics of the benchmark. Our main contributions are as follows:
\begin{itemize}[leftmargin=0cm, itemindent=0 cm]
\item We implement most of existing DSE algorithms in the literature, including seven heuristic techniques and three state-of-the-art (SOTA) RL methods, and find that no single DSE algorithm is universally optimal through extensive experiments.
	
\item We redefine the research focus of DSE by integrating algorithm selection to provide recommendations for individual benchmarks, thereby mitigating the constraints imposed by the No-Free-Lunch theorem.
	
\item To overcome the challenge of limited training data, we develop a hybrid approach that integrates supervised learning with reinforcement learning, enabling high-precision algorithm recommendation.

\item Experimental results demonstrate that our proposed SoberDSE framework achieves highly accurate algorithm recommendations, delivering an average performance improvement of 81.06\% over heuristic-based DSE algorithms and 68.80\% over learning-based DSE methods.
\end{itemize}

\section{RELATED WORKS}
\textbf{Heuristic-based Design Space Exploration}. Heuristic-based design space exploration has been demonstrated to effectively solve multi-objective optimization problems. Previous work utilizing heuristic algorithms such as NSGA-II \cite{Schafer2017}, SA \cite{Schafer2009}, and ACO \cite{Schafer2016} has achieved efficient design space exploration. To address the tendency of traditional heuristics to become trapped in local optima, Lattice-Traversing \cite{Ferretti2018} integrates a Pareto-neighbor-based selection strategy with a random selection strategy, effectively balancing exploration and exploitation. To overcome the issues of slow convergence and hyperparameter sensitivity in conventional heuristics, HGBO-DSE \cite{Kuang2023} employs Bayesian optimization to effectively resolve these challenges. Furthermore, MOEDA \cite{Yao2025} decomposes the design space exploration problem into subproblems and solves each using multi-objective evolutionary algorithms, thereby identifying a superior Pareto front.

\textbf{RL-based Design Space Exploration}. The performance of heuristic-based design space exploration is constrained by the algorithms' inability to adapt to specific benchmark features and their reliance on rigid, non-adaptive search strategies. In contrast, IRONMAN-PRO \cite{Wu2023} combines the Actor–Critic (AC) and Policy Gradient (PG) to propose the RLMD approach, which adaptively leverages benchmark features and employs flexible exploration strategies for design space exploration. This enables RLMD to discover superior Pareto fronts compared to conventional heuristic-based methods. QL-MOEA \cite{Qian2025} enhances evolutionary algorithms by addressing their inherent limitation in autonomously selecting exploration strategies. It intelligently balances the exploration-exploitation trade-off during the search process, enabling a more thorough and efficient exploration of the design space. 
\section{PRELIMINARY}
\subsection{Algorithm Selection} \label{3.1}
The algorithm selection philosophy, widely applied in multi-objective optimization, suggests that complex algorithmic design is not the only approach to further optimization. Recommending a potentially optimal algorithm based on the characteristics of a given problem instance is evidently more efficient. For a multi-objective optimization problem $\mathcal{P}$, an algorithm selection framework can be formulated as follows \cite{Kerschke2019}: given a set $\mathcal{I}$ of instances of problem $\mathcal{P}$, a set $\mathcal{A} = \left\lbrace A_{1},\ldots,A_{n} \right\rbrace$ of candidate algorithms, and a target optimization metric $m: \mathcal{A} \times \mathcal{I}$ that measures the performance of
any algorithm $A_i \in \mathcal{A}$ on instance set $\mathcal{I}$, the goal is to construct a algorithm selector $\mathcal{S}$ that maps each problem instance $j \in \mathcal{I}$ to an algorithm $\mathcal{S}(j) \in \mathcal{A}$ such that the overall performance of $\mathcal{S}$ on $\mathcal{I}$ is optimized according to metric $m$.

\subsection{Problem Formulation}\label{3.2}
In this work, we propose to leverage the algorithm selection philosophy to identify high-performing DSE algorithms tailored to specific benchmarks. Accordingly, we frame the problem in terms of the following two subproblems:

\noindent
\textbf{\textit{Problem 1: Build the Algorithmic Selection Model.}} For an FPGA accelerator kernel (benchmark) $B$ described using a high-level C-like language, its design space is denoted as $I$. This scenario involves a candidate algorithm set $\mathbf{A}$ for DSE, an optimization objective ADRS, and a pre-trained Quality of Result (QoR) prediction model $PM$. For the optimization objective ADRS, we leverage candidate algorithms from set $\mathbf{A}$ to explore the benchmark $B$. The design configurations $dc \in I$ discovered during this exploration are then evaluated using a pre-trained model $P_m$, which ultimately yields the performance result $\mathcal{H}(B)$ for each algorithm $\mathbf{A}_{k} \in \mathbf{A}$ on the benchmark $B$. The objective is to find an algorithmic recommendation function $\mathcal{F}$ that can approximate $\mathcal{H}(B)_{min}$: $min(\mathcal{H}(B))$ for any given benchmark $B$ with any  design space $I$:
\begin{equation}
	\min\limits_{\mathcal{F}} \left(Loss(\mathcal{H}(B)',\mathcal{H}(B)_{min})\right) 
\end{equation}
where $\mathcal{H}(B)'$ is the approximate value generated by function $\mathcal{F}$.

\noindent
\textbf{\textit{Problem 2: Identify the Pareto frontier Configurations.}} For the benchmarks $B$, design space $I$ and $\mathcal{F}$ defined above, $\mathcal{F}$ recommends a potentially optimal algorithm $\mathbf{A}_{best} \in \mathbf{A}$ based on the characteristics of $B$ to search for the Pareto frontier configurations $Pf$. The definition of Pareto frontier $Pf$ is as follows:
\begin{equation}
	a(d_{Pf}) \leq a(d_{I})\  and\  l(d_{Pf}) \leq l(d_{I})
	\label{4}
\end{equation}
where $d_{Pf}\in Pf$, $d_{I} \in I$ and $Pf \subseteq I$. The function $\textit{a(·)}$ and $\textit{l(·)}$ represent the resource utilization and the latency of $d_{Pf}$ and $d_{I}$.

\begin{figure*}[t]
	\centering
	\includegraphics[width=\linewidth]{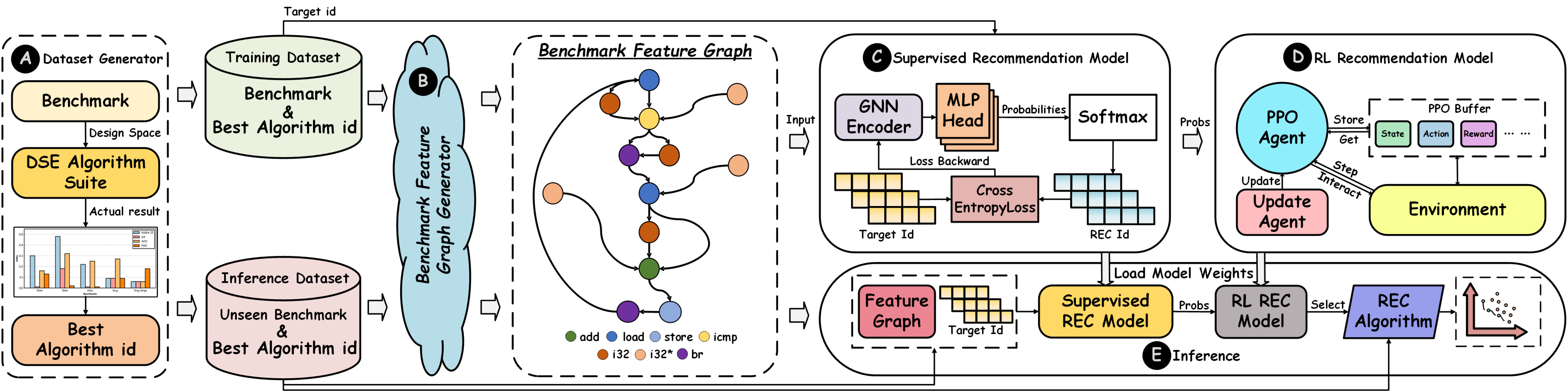}
	\caption{The Workflow of Our Proposed \textit{\textbf{SoberDSE}}.}
	\label{wf}
\end{figure*}

\section{METHODOLOGY}
As illustrated in Figure \ref{wf}, our proposed SoberDSE framework begins with the generation of training and inference data via the Dataset Generator (\circled{A}). We first implement a DSE algorithm suite comprising seven heuristic algorithms and three reinforcement learning methods. Benchmarks are divided into training and inference sets, and the performance of all DSE algorithms on each benchmark is recorded to identify the optimal algorithm for each benchmark. The Benchmark Feature Graph Generator (\circled{B}) then constructs feature graphs for each benchmark. The supervised recommendation model (\circled{C}) is initially trained and used to initialize the agent’s state in the RL recommendation model (\circled{D}) by estimating algorithm selection probabilities. The agent interacts with the environment based on ground-truth optimal algorithm IDs, learning from experiences stored in a buffer. During Inference (\circled{E}), SoberDSE is evaluated on the inference dataset generated by the Dataset Generator (\circled{A}). The pre-trained supervised and RL recommendation models (\circled{B} and \circled{C}) collaborate to suggest algorithms, which are executed on corresponding benchmarks to identify improved Pareto fronts. 

\begin{table}
	\setlength{\tabcolsep}{1pt}
	\renewcommand{\arraystretch}{1.2}
	\centering
	\footnotesize
	\caption{The Training Benchmark Used for DSE Algorithm Recommendation.}
	\begin{tabular}{l|c|c}
		\toprule
		\bf{Benchmarks} & \bf{Functions} & \bf{|DS|} \\
		\midrule
		\multicolumn{3}{c}{\textit{\textbf{MachSuite}} \cite{Reagen2014}} \\
		\hline
		gemm-blocked & O(n3) algorithm for dense matrix multiplication. & 2,314 \\
		\hline
		gemm-ncubed & Blocked version of matrix multiplication. & 7,152 \\
		\hline
		\multicolumn{3}{c}{\textit{\textbf{Polyhedral}} \cite{Yuki2010}} \\ 
		\hline
		atax & Matrix Transpose and Vector Multiplication. & 2,644 \\
		\hline 
		bicg & BiCG Sub Kernel of BiCGStab Linear Solver. & 1054 \\
		\hline
		bicg-large & BiCG Sub Kernel of BiCGStab Linear Solver. & 2,975 \\
		\hline
		covariance & Covariance Computation.& 970,967,172 \\
		\hline
		fdtd-2d & 2-D Finite Different Time Domain Kernel. & 18,329,692,529 \\
		\hline
		fdtd-2d-large & 2-D Finite Different Time Domain Kernel. & 70,303,226,437\\
		\hline
		gemm-p & Matrix-multiply.& 409,905 \\
		\hline
		gemm-p-large & Matrix-multiply. & 690,641 \\
		\hline
		gemver & Vector Multiplication and Matrix Addition. & 100,977,795,072 \\
		\hline
		gesummv & Scalar, Vector and Matrix Multiplication. & 1,543 \\
		\hline
		heat-3d & Heat equation over 3D data domain. & 71,511 \\
		\hline
		jacobi-1d & 1-D Jacobi stencil computation. & 2,871 \\
        \hline
		mvt-medium & Matrix Vector Product and Transpose. & 933,156 \\
        \hline
		syr2k & Symmetric rank-2k update. & 340,234 \\
        \hline
		symm & Symmetric matrix-multiply. & 50,265 \\
		\hline
		trmm & Triangular matrix-multiply. & 43,463 \\
		\hline
		trmm-opt & Triangular matrix-multiply. & 50,265 \\
		\hline
		2mm & 2 Matrix Multiplications & 492,787,501 \\
		\hline
		3mm &  3 Matrix Multiplications & 17,658,216,447,701 \\
		\bottomrule
	\end{tabular}
	\label{tb}
\end{table}

\subsection{Dataset Generator} \label{4.1}
To construct the dataset for DSE algorithm recommendation, as shown in Part (\circled{A}) of Figure \ref{wf}, we first implement a DSE algorithm suite consisting of seven heuristic algorithms (NSGA-II, SA, ACO, PSO, Lattice, HGBO-DSE, and MOEDA) and three RL algorithms (AC, PG, and QL-MOEA). These algorithms are executed on twenty benchmarks listed in Table \ref{tb}, with the explored design configurations evaluated using a pre-trained ECoGNN \cite{xu2025} model for QoR prediction. The benchmarks used for training ECoGNN are entirely separate from those employed for training the DSE algorithm recommendation model. To evaluate the performance of DSE algorithms, we adopt ADRS as the evaluation metric:
\begin{equation}
	ADRS(\Gamma,\Omega) = \cfrac{1}{|\Gamma|}\ \underset{\lambda \in \Gamma}{\sum}\ \underset{\mu \in \Omega}{\min}f\left(\lambda, \mu \right)
	\label{9}
\end{equation}
where $\Gamma$ represents the reference Pareto-optimal set, $\Omega$ denotes the approximate Pareto-optimal set, and the function $f$ computes the distance between $\lambda$ and $\mu$. The average performance of the ten algorithms in the DSE algorithm suite is used as the reference set for calculating the ADRS metric. Finally, we identify the algorithm ID achieving optimal performance (lowest ADRS value) for each benchmark, which serves as the training and inference target for the subsequent recommendation model.

\begin{figure}[t]
	\centering
	\includegraphics[width=\linewidth]{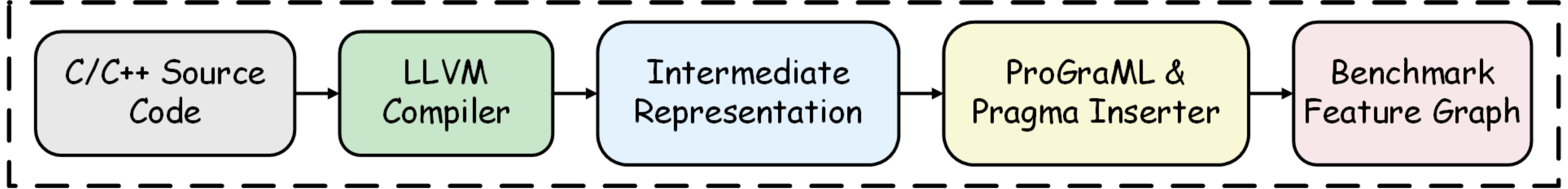}
	\caption{Benchmark Feature Graph Generation Process.}
	\label{BFGG}
\end{figure}

\subsection{Benchmark Feature Graph Generator}
As discussed in Sections \ref{3.1} and \ref{3.2}, while the recommendation model updates its parameters and biases based on training data to select algorithm IDs, the representation of benchmarks typically in high-level languages makes it challenging to fully capture their characteristic features, which is crucial for the accuracy of the recommendation algorithm. To address this issue, we propose the Benchmark Features Graph Generator (BFGG) (\circled{B}), which represent the benchmark in the form of feature graph (BFG) by using the method of generating CDFG. As illustrated in Figure \ref{BFGG}, the source code from benchmark is first converted to an intermediate representation (IR) using LLVM, followed by the generation of a BFG through ProGraML \cite{Chris2021}, which incorporates HLS directives for enhanced semantic representation. Concurrently, following the GNN-DSE \cite{Sohrabizadeh2022}, pragma inserter constructs icmp nodes to augment the BFG with pragma-specific information.

\subsection{Collaborative Recommendation Model}
Due to the reconfigurable hardware nature of FPGA accelerator kernels (benchmarks), the number of real-world benchmarks is inherently limited, which consequently restricts the number of benchmarks available for our use. As a result, the data samples obtained from the Dataset Generator (\circled{A}) are constrained in quantity. Treating algorithm recommendation solely as a classification problem using supervised learning may lead to overfitting and poor generalization. An alternative approach involves reinforcement learning, where an agent continuously interacts with the environment to recommend optimal algorithms. However, under limited sample constraints, this method still suffers from low sample efficiency and local optima convergence. To address these issues, we propose a collaborative recommendation paradigm where algorithm IDs generated by a pre-trained supervised model initialize the RL agent's state, effectively mitigating the aforementioned limitations.

As shown in Supervised Recommendation Model (\circled{C}) of Figure \ref{wf}, we first encode the BFG provided in BFGG (\circled{B}) using a pre-trained GNN encoder \cite{xu2025} to obtain the graph representation $h_{\mathcal{G}}$. The GNN encoder employed here corresponds to the encoder component of the QoR prediction model. This encoding process can be formulated as:
\begin{equation}
	 h_{\mathcal{G}} = GNN_{encoder}(BFG)
\end{equation}
Then, we set up an MLP head to decode $h_{\mathcal{G}}$ and add a softmax layer to output the recommendation probability for each algorithm. This process can be formalized as:
\begin{equation}
	p(A|BFG) = \text{softmax}(W_2 \cdot \text{ReLU}(W_1 h_{\mathcal{G}} + b_1) + b_2)
\end{equation}
where $W_1, W_2, b_1$ and $b_2$ are the parameter of MLP head, $A$ is the set of DSE algorithms. We employ the Cross-Entropy loss \cite{Mao2023} to compute the discrepancy between the algorithm ID recommended by the supervised model and the ground-truth best-performing algorithm ID during training.

After a brief training period, we save the parameters of the supervised model and transmit its final output recommendation probabilities $p_{sup}$ to the RL recommendation model. Since the limited sample size, we employ the PPO algorithm \cite{Schulman2017} as our reinforcement learning approach due to its high sample efficiency and training stability. As shown in RL Recommendation Model (\circled{D}) of Figure \ref{wf}, The PPO agent operates in the environment where the state representation $s_t = [h_G; p_{sup}]$ combines graph embeddings $h_{\mathcal{G}}$ from the GNN encoder and prior probabilities $p_{sup}$ from the supervised model. To enable interaction with the environment, the PPO agent incorporates two trainable neural networks: an actor network and a critic network. The actor network selects algorithm $a_t \in A$ based on the current state $s_t$:
\begin{equation}
	\pi_\theta(a_t|s_t) = \text{softmax}(\text{MLP}_a(s_t))
\end{equation}
where $MLP_a$ is used to approximate the probability distribution of algorithm recommendations $p(a_t|s_t)$. The critic network estimates state values for advantage calculation:
\begin{equation}
	V_\phi(s_t) = \text{MLP}_c(h_G)
\end{equation}
where $MLP_a$ is used to calculate state values. The environment provides rewards $r_t$ based on performance deviation:
\begin{equation}
	r_t = \frac{|m^{(a_t)} - \bar{m}|}{\bar{m}}
\end{equation}
where $m^{(a_t)}$ represents the ADRS value of algorithm $a_t$ and $\bar{m}$ denotes the ADRS value of the best-performing algorithm on the corresponding benchmark. The PPO buffer serves as a critical component in the training pipeline by facilitating the interaction between the agent and environment while enabling stable policy updates. 

During each episode, the agent interacts with the environment by selecting actions $a_t$ based on the current state $s_t$, receiving rewards $r_t$, and observing state transitions. These experiences ($s_t, a_t, r_t, s_{t+1}, log(\pi_\theta(a_t|s_t)), V_\phi(s_t)$) are stored in the PPO buffer. After episode completion, the buffer provides the complete trajectory data for policy optimization. The update process involves computing advantages using Generalized advantage estimation (GAE) \cite{Schulman2015}:
\begin{equation}
	A_t = \sum_{l=0}^{\infty} (\gamma\lambda)^l \delta_{t+l}
\end{equation}
where $\delta_t = r_t + \gamma V(s_{t+1}) - V(s_t)$, $\gamma,\lambda$ are hyperparameters of GAE. The agent is then updated by minimizing the PPO objective:
\begin{equation}
	\mathcal{L}_{PPO} = \mathcal{L}_{policy} + c_1 \mathcal{L}_{value} - c_2 \mathcal{L}_{entropy}
\end{equation}
where $\mathcal{L}_{policy}$ is the clipped policy gradient loss that encourages policy improvement while constraining the update size to prevent drastic policy changes, $\mathcal{L}_{value}$ is the value function loss that trains the critic network to accurately estimate state values, $\mathcal{L}_{entropy}$ is the entropy bonus that encourages exploration by increasing the randomness of the policy and $c_1, c_2$ are hyperparameters. 

By training the PPO agent and saving its parameters, we evaluate the proposed recommendationl
 model as shown in the Inference (\circled{E}) section of Figure \ref{wf}. First, the benchmark features graph from the inference dataset is input to the pre-trained supervised recommendation model to generate an initial algorithm recommendation probability. The RL recommendation model then integrates this probability to suggest the potentially optimal DSE algorithm and returns the Pareto front obtained by that algorithm on the corresponding benchmark. 
\begin{figure}[t]
	\centering
	\includegraphics[width=0.90\linewidth]{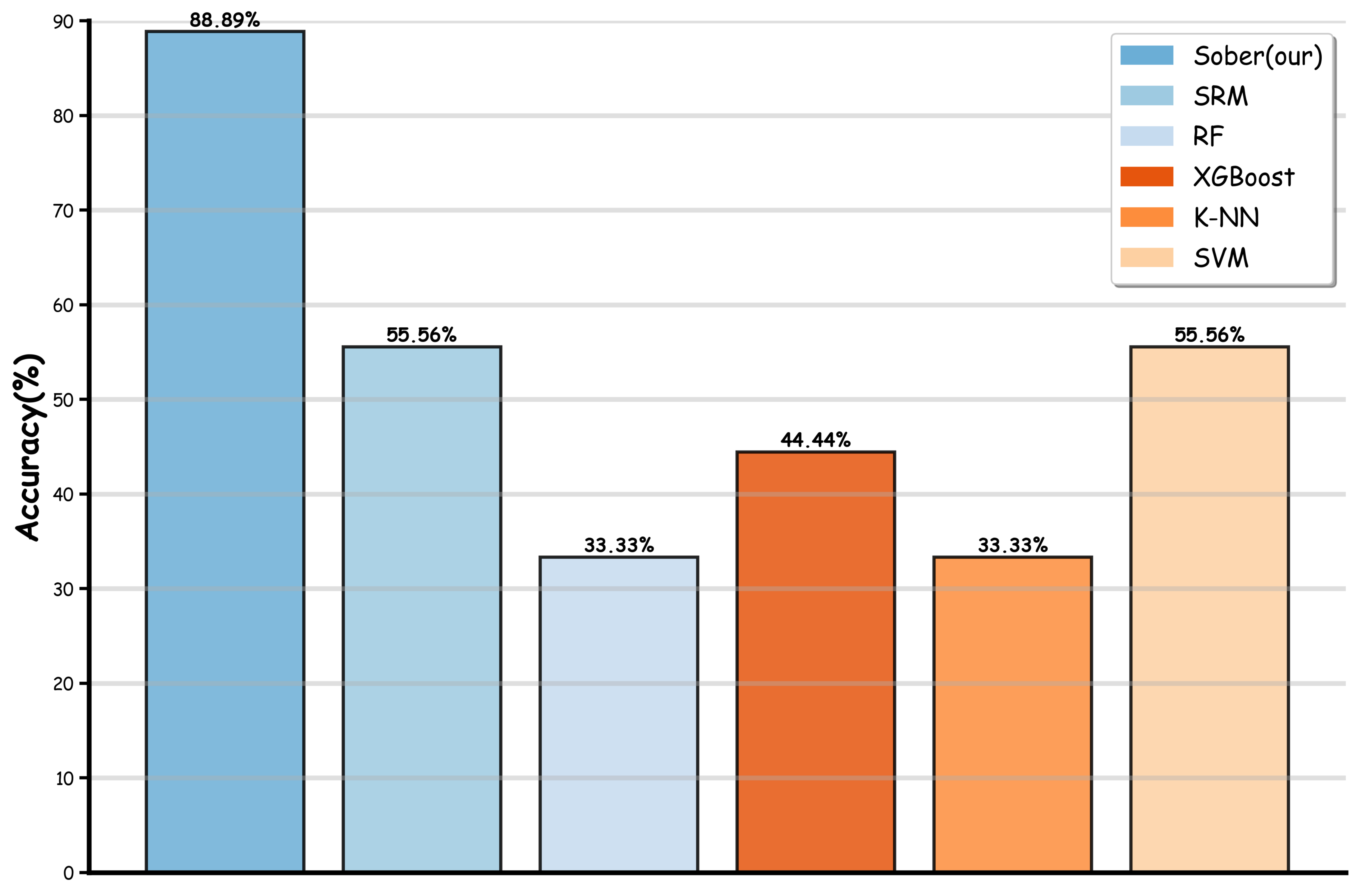}
	\caption{Comparing Recommendation Accuracy of \textit{Sober}, SFM and Baseline Models}
	\label{rr}
     \vspace{-1mm}
\end{figure}

\begin{table}
	\setlength{\tabcolsep}{1.2pt}
	\renewcommand{\arraystretch}{1.4}
	\centering
	\footnotesize
	\caption{The Inference Benchmark Used for DSE Algorithm Recommendation.}
	\begin{tabular}{l|c|c}
		\toprule
		\bf{Benchmarks} & \bf{Functions} & \bf{|DS|} \\
		\midrule
		\multicolumn{3}{c}{\textit{\textbf{MachSuite}} \cite{Reagen2014}} \\
		\hline
		nw & A dynamic programming algorithm. & 15,288 \\
		\hline
		stencil & A two-dimensional stencil computation. & 7,591 \\
		\hline
		\multicolumn{3}{c}{\textit{\textbf{Polyhedral}} \cite{Yuki2010}} \\ 
		\hline
		atax-medium & Matrix Transpose and Vector Multiplication. & 9,454 \\
		\hline 
		bicg-medium & BiCG Sub Kernel of BiCGStab Linear Solver. & 19,536 \\
		\hline
		correlation & Correlation Computation.& 513,641,633,988 \\
		\hline
		jacobi-2d & 2-D Jacobi stencil computation. & 7,609,187 \\
		\hline
		mvt& Matrix Vector Product and Transpose.  & 2,385,796\\
		\hline
		symm-opt & Symmetric matrix-multiply. & 730,665 \\
		\hline
		syrk & Symmetric rank-k update. & 340,234 \\
		\bottomrule
	\end{tabular}
	\label{Ib}
     \vspace{-3mm}
\end{table}

\section{EXPERIMENTAL EVALUATIONS}
\subsection{Experimental Setup}
As outlined in Section \ref{4.1}, we construct datasets for training and evaluating the proposed SoberDSE framework. The training set comprises 20 benchmarks listed in Table \ref{tb}. As shown in Table \ref{Ib}, the inference dataset comprises 9 benchmarks sourced from MachSuite \cite{Reagen2014} and Polyhedral \cite{Yuki2010}, exhibiting significant diversity in design space scale including the hundred-billion-level benchmark (\textit{correlation}). The benchmarks for algorithm recommendation are completely separate from those used for QoR prediction model training. Additionally, the main methods we compare include seven heuristic algorithms (NSGA-II \cite{Schafer2017}, SA \cite{Schafer2009}, ACO \cite{Schafer2016}, PSO \cite{Zhang2015}, Lattice \cite{Ferretti2018}, HGBO-DSE \cite{Kuang2023}, and MOEDA \cite{Yao2025}) and three reinforcement learning algorithms (IRONMAN-PRO \cite{Wu2023} with AC and PG, and QL-MOEA \cite{Qian2025}). The supervised model employs a hidden layer dimension of 256, a learning rate of 0.0003, and 250 training epochs, while the reinforcement learning model uses a hidden layer dimension of 256, a learning rate of 0.0005 and 1000 training epochs. All experiments are executed on the Xilinx Alveo U200 FPGA with a frequency of 250MHz. All DSE algorithms are implemented in Python, with the SoberDSE framework developed and trained using PyTorch.

\subsection{Recommendation Accuracy Analysis} \label{5.2}
We utilize the trained Sober((\circled{C}) and (\circled{D})) to perform algorithm recommendations on the inference benchmarks following the Inference (\circled{E}) workflow illustrated in Figure \ref{wf}. To validate the Recommendation accuracy (percentage of correct recommendations relative to the total number of samples) of our proposed algorithm, we conduct comparative evaluations of Sober and against classification methods including Random Forest (RF) \cite{Breiman2001}, XGBoost \cite{Chen2016}, K-NN \cite{Cover1967}, and SVM \cite{Cortes1995}. Furthermore, to validate the importance of the PPO component, we conduct experiments using only the Supervised recommendation model (SFM) for algorithm recommendation and recorded the results. As shown in Figure \ref{rr}, our proposed Sober recommendation model demonstrates significantly more accurate algorithm recommendations compared to traditional classification models, achieving an average improvement of 35.57\%. In contrast, the SFM without the PPO component shows comparable accuracy to baseline models, indicating the limited performance of supervised models on small-sample problems and their susceptibility to overfitting. Sober effectively addresses the overfitting issue in supervised learning by integrating reinforcement learning. Additionally, by initializing the RL recommendation model's state with probabilities from the supervised model, Sober enhances the sample efficiency of the RL recommendation process.

\begin{table*}
	\setlength{\tabcolsep}{5pt}
	\renewcommand{\arraystretch}{1.0}
	\normalsize
	\centering
	\caption{Comparison of ADRS Results between SoberDSE and SOTA DSE Heuristic Algorithms.}
	\begin{tabular}{l|c|c|c|c|c|c|c|c}
		\toprule
		\bf{Benchmarks} & \bf{NSGA-II\cite{Schafer2017}} & \bf{SA\cite{Schafer2009}} & \bf{ACO\cite{Schafer2016}} & \bf{PSO\cite{Zhang2015}} & \bf{Lattice\cite{Ferretti2018}} & \bf{HGBO-DSE\cite{Kuang2023}} & \bf{MOEDA\cite{Yao2025}} & \bf{SoberDSE(Our Work)}\\
		\midrule
		nw & 0.0275 & 0.0275 & 0.0596 & 0.0275 & \textbf{0.0019} & 0.0122 & 0.0282 & 0.0046\\
		\hline
		stencil & 0.0896 & 0.0896 & 0.1870 & 0.2167 & 0.3105 & 0.4714 & 0.1551 & \textbf{0.0189}\\
		\hline
		atax-medium & 0.0638 & 0.0219 & 0.0564 & 0.0219 & 0.0715 & 0.0328 & 0.0913 & \textbf{0.0171}\\
		\hline 
		bicg-medium & 0.1043 & 0.1043 & 0.0650 & 0.1778 & 0.1043 & 0.0401 & 0.1125 & \textbf{0.0256}\\
		\hline
		correlation & 0.1726 & 0.0608 & 0.2692 & 0.1287 & 0.0185 & \textbf{0.0142} & 0.1419 & 0.0145\\
		\hline
		jacobi-2d & 0.2215 & 0.1245 & 0.2750 & 0.2200 & 0.2537 & \textbf{0.0153} & 0.1023 & 0.0188\\
		\hline
		mvt & 0.3017 & 0.1787 & 0.1924 & 0.0595 & 0.0337 & 0.0688 & 0.1151 & \textbf{0.0258}\\
		\hline
		symm-opt & 0.1555 & 0.0649 & 0.0649 & 0.0526 & 0.0274 & 0.0334 & 0.0777 & \textbf{0.0205}\\
		\hline
		syrk & 0.1144 & 0.0833 & 0.1885 & 0.0516 & \textbf{0.0072} & 0.0358 & 0.0991 & 0.0091\\
		\hline
		Avg & 0.1390 & 0.0839 & 0.1509 & 0.1063 & 0.0921 & 0.0804 & 0.1026 & \textbf{0.0172}\\
		\hline
		\multicolumn{5}{c|}{\bf{SoberDSE Improv. over Lattice}}&\multicolumn{4}{c}{\bf{81.32\%}} \\
		\multicolumn{5}{c|}{\bf{SoberDSE Improv. over HGBO-DSE}}&\multicolumn{4}{c}{\bf{78.61\%}} \\
		\multicolumn{5}{c|}{\bf{SoberDSE Improv. over MOEDA}}&\multicolumn{4}{c}{\bf{83.24\%}} \\
		\bottomrule
	\end{tabular}
	\label{DSET}
    \vspace{-4mm}
\end{table*}

\begin{figure}[t]
	\centering
	\includegraphics[width=0.90\linewidth]{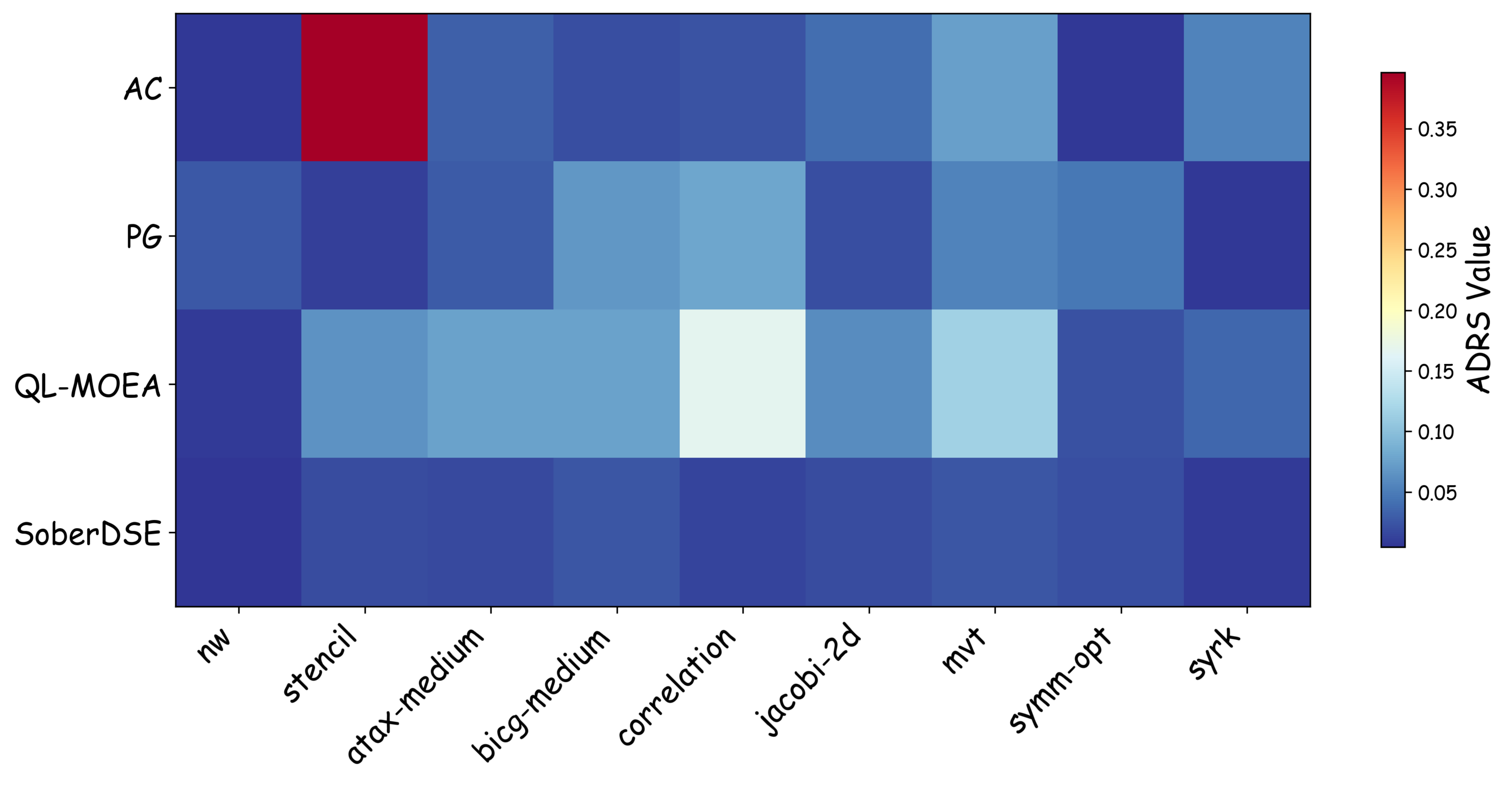}
	\caption{Comparison of ADRS Results between \textit{SoberDSE} and SOTA DSE RL Algorithms.}
	\label{hm}
    \vspace{-3mm}
\end{figure}

\subsection{DSE Performance Analysis} \label{5.3}
We evaluate the algorithms recommended by Sober ((\circled{C}) and (\circled{D})) on the benchmarks listed in Table \ref{Ib} using the ADRS metric, comparing primarily with heuristic techniques including NSGA-II, SA, ACO, PSO, Lattice, HGBO-DSE, and MOEDA. Since our approach involves algorithm recommendation followed by executing the recommended algorithm on the corresponding benchmark, the results may coincide with one of the comparative algorithms. However, due to the stochastic nature of exploratory algorithms, different runs may produce non-deterministic outcomes. Experimental results in Table \ref{DSET} show that SoberDSE achieves the lowest average ADRS, outperforming the SOTA DSE heuristic algorithms Lattice, HGBO-DSE, and MOEDA by 81.32\%, 78.61\%, and 83.24\%, respectively. This demonstrates that the algorithm selection strategy effectively mitigates the limitations of the NFL theorem, enabling SoberDSE to deliver superior overall performance across all benchmarks by recommending the potentially optimal algorithm for each case. Additionally, we also compare with SOTA DSE RL algorithms including IRONMAN-PRO (AC and PG) and QL-MOEA. As shown in Figure \ref{hm}, darker blue shades indicate better algorithm performance on corresponding benchmarks. Experimental results demonstrate that our proposed SoberDSE outperforms SOTA DSE RL algorithms, achieving the best average performance across all benchmarks. Specifically, SoberDSE improves over AC by 76.31\%, PG by 54.81\%, and QL-MOEA by 75.29\%. The stable and superior exploration capability of SoberDSE confirms the feasibility of using algorithm selection to address DSE optimization challenges.

\begin{figure}[t]
	\centering
	\includegraphics[width=\linewidth]{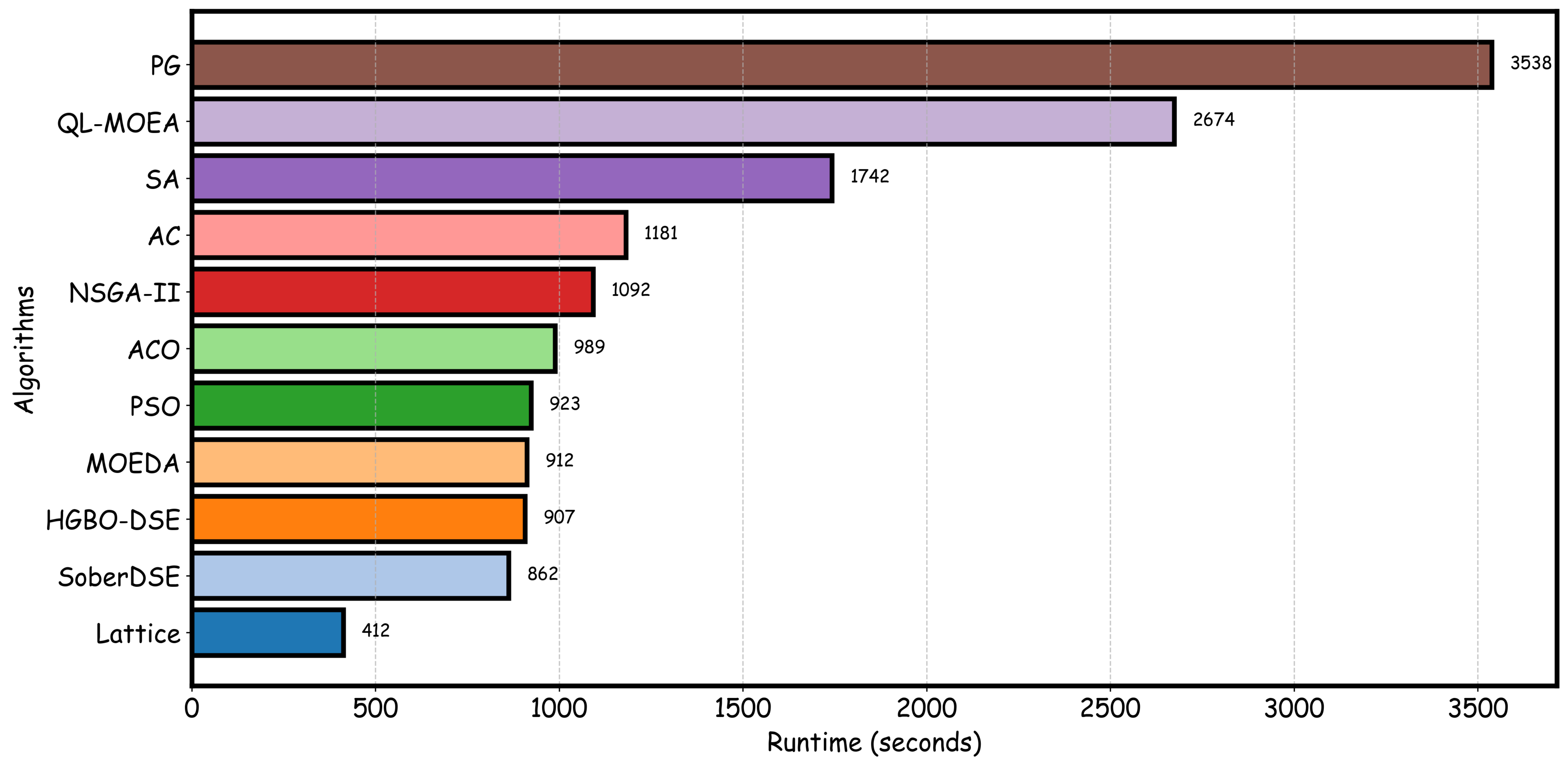}
	\caption{Comparison of Runtime between \textit{SoberDSE} and 10 Baseline Algorithms. }
	\label{rt}
    \vspace{-6mm}
\end{figure}

\subsection{Runtime Analysis} \label{5.4}
In the experiments, we also compare the runtime of SoberDSE and all algorithms. As shown in Figure \ref{rt}, SoberDSE demonstrates significant runtime advantages on the 9 benchmarks mentioned in Table \ref{Ib}, achieving an overall runtime of 862s. This outperforms most other algorithms, including NSGA-II (1092s), SA (1742s), ACO (989s), PSO (923s), PG (3538s), and QL-MOEA (2674s). While remaining competitive with MOEDA (912s) and HGBO-DSE (907s), it is slightly slower than Lattice (412s). This represents a reduction of up to 75.63\% in runtime compared to these DSE algorithms. This demonstrates that SoberDSE achieves superior performance compared to other baseline algorithms while requiring relatively less computational time.

In summary, based on the evaluation in Sections \ref{5.2}-\ref{5.4}, our proposed SoberDSE method achieves a higher recommendation accuracy than conventional classification models and surpasses SOTA DSE algorithms with relatively low time consumption. These results indicate that addressing DSE challenges need not rely solely on designing new algorithms, but can also be accomplished by recommending suitable algorithms from existing suites based on benchmark characteristics, thereby achieving overall optimal performance. This paradigm shift offers a powerful alternative to designing new algorithms from scratch.

\section{CONCLUSION}

We present SoberDSE, an automated algorithm selection framework that extracts benchmark features and selects optimal algorithms from a suite of DSE tools. To enable efficient selection with limited samples, we introduce a hybrid recommendation model combining supervised and reinforcement learning, which significantly outperforms conventional classifiers. Under limited-sample conditions, SoberDSE achieves superior performance compared to state-of-the-art heuristic and reinforcement learning models, but in a fraction of the time. This work offers a novel perspective on DSE optimization challenges. Future research will expand the algorithm suite and train SoberDSE on additional benchmarks to further enhance its performance.

\nocite{*}
\bibliographystyle{ACM-Reference-Format}
\bibliography{acmart}

\end{document}